\documentclass[%
 aip,
 amsmath,amssymb,
 reprint,%
]{revtex4-1}

\usepackage{graphicx}
\usepackage{dcolumn}
\usepackage{bm}

\usepackage[utf8]{inputenc}
\usepackage[T1]{fontenc}
\usepackage{mathptmx}
\usepackage{etoolbox}
\usepackage{xcolor}
\usepackage{multirow}

\makeatletter
\def\@email#1#2{%
 \endgroup
 \patchcmd{\titleblock@produce}
  {\frontmatter@RRAPformat}
  {\frontmatter@RRAPformat{\produce@RRAP{*#1\href{mailto:#2}{#2}}}\frontmatter@RRAPformat}
  {}{}
}%
\makeatother
\begin{document}

\preprint{AIP/123-QED}

\title[The vertical velocity skewness in the atmospheric boundary layer without buoyancy and Coriolis effects]{The vertical velocity skewness in the atmospheric boundary layer without buoyancy and Coriolis effects}
\author{Elia Buono}
\affiliation{Department of Civil and Environmental Engineering, Duke University, Durham, North Carolina, USA}
\affiliation{Dipartimento di Ingegneria dell'Ambiente, del Territorio e delle Infrastrutture, Politecnico di Torino, Torino, Italia}
\email{elia.buono@polito.it}
 
\author{Gabriel Katul}%
\affiliation{Department of Civil and Environmental Engineering, Duke University, Durham, North Carolina, USA}%

\author{Michael Heisel}
\affiliation{School of Civil Engineering, University of Sydney, Sydney, Australia}%

\author{Davide Poggi}
\affiliation{Dipartimento di Ingegneria dell'Ambiente, del Territorio e delle Infrastrutture, Politecnico di Torino, Torino, Italia}%

\author{Cosimo Peruzzi}
\affiliation{Area for Hydrology, Hydrodynamics, Hydromorphology and Freshwater Ecology (BIO-ACAS), Italian Institute for Environmental Protection and Research (ISPRA), Rome, Italy}%

\author{Davide Vettori}
\affiliation{Dipartimento di Ingegneria dell'Ambiente, del Territorio e delle Infrastrutture, Politecnico di Torino, Torino, Italia}

\author{Costantino Manes}
\affiliation{Dipartimento di Ingegneria dell'Ambiente, del Territorio e delle Infrastrutture, Politecnico di Torino, Torino, Italia}%

\date{\today}

\begin{abstract}
One of the main statistical features of near-neutral atmospheric boundary layer (ABL) turbulence is the positive vertical velocity skewness $Sk_w$ above the roughness sublayer or the buffer region in smooth-walls.  The $Sk_w$ variations are receiving renewed interest in many climate-related parameterizations of the ABL given their significance to cloud formation and to testing sub-grid schemes for Large Eddy Simulations (LES).  The vertical variations of $Sk_w$ are explored here using high Reynolds number wind tunnel and flume experiments collected above smooth, rough, and permeable-walls in the absence of buoyancy and Coriolis effects.  These laboratory experiments form a necessary starting point to probe the canonical structure of $Sk_w$ as they deal with a key limiting case (i.e. near-neutral conditions) that has received much less attention compared to its convective counterpart in atmospheric turbulence studies.  Diagnostic models based on cumulant expansions, realizability constraints, and the now-popular constant mass flux approach routinely employed in the convective boundary layer as well as prognostic models based on third-order budgets are used to explain variations in $Sk_w$ for the idealized laboratory conditions.  The failure of flux-gradient relations to model $Sk_w$ from the gradients of the vertical velocity variance $\sigma_w^2$ are explained and corrections based on models of energy transport offered. Novel links between the diagnostic and prognostic models are also featured, especially for the inertial term in the third order budget of the vertical velocity fluctuation.  The co-spectral properties of $w'/\sigma_w$ versus $w'^2/\sigma_w^2$ are also presented for the first time to assess the dominant scales governing $Sk_w$ in the inner and outer layers, where $w'$ is the fluctuating vertical velocity and $\sigma_w$ is the vertical velocity standard deviation.   

\end{abstract}
\maketitle
\section{Introduction}
\label{sec:introduction}
Turbulent motion is responsible for much of the transport of heat and water vapor within the planetary boundary layer \citep{peixoto1984physics}.  This transport determines the distribution of temperature, winds, cloud formation, and precipitation \citep{trenberth1992climate}.  For this reason, it is often stated that life on Earth as we know it would not be possible without turbulence \citep{sreenivasan1999fluid}.  Climate is routinely viewed as a long-term integrator of weather (measured in decades) - an assertion put forth by the Nobel laureate K. Hasselmann \citep{hasselman1976stochastic}.  In Hasselmann's representation, the coupled ocean-atmosphere-cryosphere-land system is decomposed into a rapid part - the “weather” system (essentially the atmosphere) and a slowly responding “climate” system (mainly the ocean, cryosphere, and land vegetation).  Weather is defined as the average state of the atmosphere determined by its temperature, atmospheric pressure, wind, humidity, precipitation, and cloud cover.  Once again, averaging of these atmospheric states is required to integrate stochastic fluctuations in the aforementioned variables - and those stochastic fluctuations are traditionally attributed to turbulence.  Turbulent time scales in the atmosphere span fraction of seconds (at the micro-scales) to an hour or so (for large and very large eddy motion), and it is the aggregate of these fluctuations that generate turbulent fluxes and changes in the mean state of atmospheric variables needed for simulating or modeling weather \citep{stull2012introduction}.  

Turbulence in the atmospheric boundary layer (ABL) has a number of well-established 'signatures' in its statistics that are deemed significant for climate and meteorological modeling, dispersion studies, wind energy generation, and a plethora of other applications pertinent to atmospheric chemistry and atmospheric composition that 'feed-back' on climate \citep{fuentes2016linking}.  The skewness of the vertical component $Sk_w$ has long been recognized as one such key feature \citep{wyngaard_2010} and frames the scope of this work.  It is the most elementary flow statistics quantifying asymmetry due to the presence of a boundary and is given by 
\begin{equation}
\label{eqn:Skw_def}
    Sk_w=\frac{\overline{w'^3}}{\sigma_w^3},
\end{equation}
where $w'$ is the instantaneous vertical (or wall-normal) velocity fluctuation, overline denotes averaging over coordinates of statistical homogeneity (e.g. time averaging in many laboratory and field experiments or space-time averaging in direct numerical simulations), and $\sigma_s=(\overline{s's'})^{1/2}$ is the standard deviation or root-mean squared value of the fluctuations of any turbulent flow variable $s$.  

In the absence of any thermal stratification or Coriolis effects, laboratory measurements of $Sk_w$ over rough turbulent boundary layers (including k-type and d-type) suggest that $Sk_w<0$ in the roughness sublayer (RSL) or the buffer region of smooth walls but switches to $Sk_w>0$ in the inertial and outer-layers \citep{nakagawa1977prediction,raupach1981conditional,heisel2020velocity}. This switch was shown to be linked to a change in the type of organized eddy motion dominating momentum transport.  Sweeping motion dominates within the roughness sublayer and ejective motion in the remaining layers of the boundary layer \citep{raupach1981conditional,nakagawa1977prediction,poggi2004effect}.  Likewise, $Sk_w$ over natural and artificial canopies with different densities follow similar expectations with $Sk_w<0$ within and just above the canopy top followed by $Sk_w>0$ for the remaining layers \citep{poggi2004effect,raupach1981conditional,shaw1987calculation}, a result consistent with other shear-driven boundary layer experiments \citep{maurizi2006dependence}. This finding indicates that $\partial Sk_w/\partial z>0$ for much of the boundary layer depth $\delta$ except in the buffer layer for smooth walls or roughness sublayer in rough or permeable walls, where $z$ is the distance from the ground or zero plane displacement for canopy flows.  

Moving onwards to the unstable atmospheric surface layer (ASL), early studies found that $Sk_w$ follows expectations from Monin-Obukhov surface layer similarity theory \citep{monin1954basic}, hereafter referred to as MOST, and is empirically given by \citep{chiba1978stability}
\begin{equation}
\label{eqn:ASL_Skw}
   Sk_w=C_{ASL}-\frac{0.6 \xi}{1.25^3 \kappa \left[ (1-15 \xi)^{-1/4}-1.8 \xi\right]},
\end{equation}
where $C_{ASL}$ is a similarity coefficient, $\xi=z/L_o$ is the atmospheric stability parameter (for unstable atmospheric conditions, $\xi<0$) , $L_o$ is the Obukhov length \citep{stull2012introduction}, and $\kappa=0.4$ is the von K\'arm\'an constant.  These relations are consistent with the benchmark Kansas experiment that empirically demonstrated \citep{wyngaard1971budgets}
\begin{equation}
\label{eqn:updraft}
  \frac{\kappa z}{u_*^3}\frac{\partial \overline{w'^3}}{\partial z} =-0.6 \xi.
\end{equation}
For near-neutral ASL conditions (i.e. $\xi=0$), these experiments suggest that  $Sk_w=C_{ASL}\approx 0.1$ (i.e. a positive integration constant arising from equation \ref{eqn:updraft}), but offer no explanation as to why. Convective boundary layers (CBLs) are also characterized by $Sk_w>0$ over their entire depth $\delta$ \citep{lemone1990}.  In fact, experiments and Large Eddy Simulations (LES) have shown that in the CBL, $Sk_w=0.5-0.6$ persists for extended regions of the CBL ($0.2<z/\delta<0.75$) as discussed elsewhere \citep{lenschow2012comparison}, though some LES results \citep{ghannam2017non} appear to exceed field experiments in the upper regions of the CBL by a factor of two \citep{lenschow2012comparison}.  In the limit of free convection (i.e. $\xi\rightarrow-\infty$), equation \ref{eqn:ASL_Skw} suggests that $Sk_w$=$C_{ASL}$+[0.6/(0.78$\times$1.8)]=0.53, which is close to the reported $Sk_w=0.5$ from aircraft measurements for much of the CBL \citep{lemone1990} as well as recent measurements in the near-convective ASL \citep{barskov2023relationships}.  Once again, such positive near-neutral $Sk_w$ limit in the absence of stratification is not well explained. 

Interest in $Sk_w$ has been proliferating in numerous applications including non-Gaussian models for dispersion \citep{baerentsen1984monte,luhar1989random,wyngaard1991transport,maurizi1999velocity}, parameterizing subgrid schemes in LES \citep{lemone1990,sullivan2011effect,moeng1990vertical}, delineating the fraction of time turbulent flows reside in updrafts ($w'>0$) versus downdrafts ($w'<0$) \citep{quintarelli1990study}, and higher-order closure modeling  of boundary layers \citep{zilitinkevich1999third,ghannam2017non,barskov2023relationships}. The latter higher-order schemes are now being implemented in climate models such as the Cloud Layers Unified By Binormals (CLUBB) \citep{huang2020assessing,bogenschutz2012unified}, among others \citep{mellor1982development}.  
Third-order closure schemes, especially for $Sk_w$, were shown to be necessary for cloud formation in different boundary layer regimes.  Some studies showed that accommodating vertical velocity probability density function (PDF) asymmetry in the ABL increased low cloud fraction by 20$\%$ - 30$\%$ in stratocumulus-to-cumulus transition regions \citep{li2022updated}.

While much attention has been devoted to $Sk_w$ within the CBL \citep{wyngaard_2010}, less attention has been paid to models of $Sk_w$ in near-neutral conditions, the focus here.  These conditions are prevalent in many planetary flow situations (e.g. air flow over ice sheets, large open water bodies, and in many occasions over land) and form a logical limit for ABL characterizations of asymmetry, especially for vertical turbulent transport (including $Sk_w$).  They also form limiting states for approaches such as CLUBB that must be recovered \citep{waterman2022examining}.  Indeed, when comparing models and LES to field experiments, there is unavoidable bias regarding  heterogeneity at the ground in field experiments.  Heterogeneity is known to have a higher impact on variances (e.g. $\sigma_w^2$) than $\overline{w'w'w'}$ \citep{moeng1990vertical,mason1989large,lemone1990}.  Thus, disagreement between LES and field experiments, even for near-neutral limits, may be due to either subgrid filtering schemes employed by the LES or simply due to the non-ideal nature of the ground surface.  This attribution deficiency underscores the need for benchmark data and theories on $Sk_w$ derived from idealized laboratory conditions at some reference stability (e.g. near-neutral), the goal here.

In the present work, we purposely distinguish between two types of models for $Sk_w$: diagnostic and prognostic. Diagnostic models derive relations between $Sk_w$ (the target variable) and other statistical moments without requiring information about the physics of turbulence.  They often approximate the PDF of a flow variables with cumulant expansion truncated at some order, usually third or fourth \citep{nakagawa1977prediction,raupach1981conditional}.  These diagnostic models can then be used to impose constraints such as the so-called realizability condition \citep{andre1976turbulence,zilitinkevich1999third} first studied in the context of locally homogeneous and isotropic turbulence \citep{millionshchikov1941theory}. Prognostic models seek to predict $Sk_w$ from lower-order moments that can be modeled from the ensemble-averaged Navier-Stokes equations (NSE) using closure schemes.  One common closure scheme links $\overline{w'^3}$ to the vertical gradients of $\overline{w'w'}$ using an eddy-viscosity coefficient ($K_t$) given as \citep{launder1975progress,mellor1982development,lumley1979computational,hanjalic2002}
\begin{equation}
\label{eqn:www_gradww}
     \overline{w'w'w'}=-K_t \frac{\partial\overline{w'w'}}{\partial z},
\end{equation}
where $z$ is the distance from the boundary or zero-plane displacement in the case of canopy flows.  Such closure remains controversial, especially in CBL and canopy flows \citep{corrsin1975limitations,katul1998investigation,poggi2004momentum,ghannam2017non}. This motivated the development of other approaches for the CBL such as the so-called large-eddy skewed turbulence advection velocity approach or the eddy-diffusivity mass flux approach \citep{abdella2000third,zilitinkevich1999third,ghannam2017non}.   In these revisions, equation \ref{eqn:www_gradww} is adjusted using an additive term that reflects large-scale transport commonly modeled using the aforementioned mass flux approach \citep{abdella2000third}.  In such a framework, the mathematical form for $\overline{w'^3}$ is given by  \citep{deardorff1972theoretical} 
\begin{equation}
\label{eqn:www_gradww2}
     \overline{w'w'w'}=-K_t \left[\frac{\partial\overline{w'w'}}{\partial z}-\gamma_w\right],
\end{equation}
where $\gamma_w$ is a transport term formed from a large-scale advection velocity and a characteristic time scale.  In the CBL, $\gamma_w$ can be related to the vertical transport of the heat flux \citep{canuto1994second,zilitinkevich1999third}, meaning that $\gamma_w\rightarrow0$ as near-neutral conditions are approached. Despite these amendments, the frustration in satisfactorily closing $\overline{w'^3}$ or $Sk_w$ remains and is captured by the statement from a leading authority \citep{zilitinkevich1999third} who concluded that "{\it Certainly, the problem of parameterization of $Sk_w$ remains. The authors should admit that they have no definitive answer at the moment.}".  

The two classes of models (diagnostic and prognostic) are explored herein using a unique family of data sets that include open channel flow and wind tunnel experiments over smooth walls, rough walls, and permeable walls across wide-ranging Reynolds numbers $Re_\delta=\delta u_*/\nu$ and measurement techniques, where $\delta$ is the boundary layer depth or, more in general, the outer length scale of the flow and $u_*=\sqrt{\tau_w/\rho}$ is the friction velocity based on the total or wall stress $\tau_w$, $\rho$ is the fluid density (air or water), and $\nu$ is the fluid kinematic viscosity.  Moreover, how constraints derived from diagnostic models can be used to offer new parameterizations for prognostic models are discussed.  As shall be seen, the present work leads to $\overline{w'^3}$ as
\begin{equation}
\label{eqn:www_gradww_gen}
     \overline{w'w'w'}=-\overbrace{K_t \frac{\partial\overline{w'w'}}{\partial z}}^{\rm local~effects}+\underbrace{\beta_L \overline {w'q'}}_{\rm large~scale~adjustment},
\end{equation}
where $q'$ is related to the instantaneous turbulent kinetic energy, and $\beta_L$ is a constant that can be derived from pressure and viscous effects on the third order budgets of $\overline{w'^3}$.  In this derivation, both a local gradient-diffusion term and a non-local transport term arise from the governing equations.  Moreover, the signature of the non-local or large scale effects appearing through $\overline{w'q'}$ are also confirmed through analysis of the co-spectrum between $w'$ and $w'^2$ of the aforementioned laboratory studies.   The experiments here will also reveal that these non-local effects are dominant over much of the boundary depth (i.e. above the roughness sublayer or the buffer region) and that they can be parameterized by a down-gradient of the turbulent kinetic energy (instead of $\partial \overline{w'^2}/\partial z$) provided an outer layer correction is accommodated in the eddy viscosity.  The $\overline{w'q'}$ has already been linked through quadrant analysis and conditional sampling to the relative importance of sweeps and ejections on momentum transport \citep{nakagawa1977prediction,raupach1981conditional}.  However, those links - sometimes termed as structural models \citep{nagano1988statistical,nagano1990structural,olccmen2006octant}-  remain diagnostic \citep{heisel2020velocity,poggi2004momentum,cava2006buoyancy}.  Thus, a key novelty of the present work is a link across all the aforementioned approaches and their testing in smooth, rough, and permeable boundaries restricted to diabatic non-rotating flows that are stationary and planar homogeneous. 

\section{Theory}
\label{sec:theory}

\subsection{Definitions}
The Cartesian coordinate system employed here sets $x=x_1$, $y=x_2$, and $z=x_3$ along the longitudinal, lateral, and vertical (wall-normal) directions, respectively, with $z=0$ being the ground or zero-plane displacement.  The instantaneous velocity components along $x$, $y$, and $z$ directions are labeled as $u=u_1$, $v=u_2$, and $w=u_3$, respectively, with $U=\overline{u}$ defining the mean velocity.  Velocity fluctuations from their time-averaged values at a point are indicated by primed quantities. 

\subsection{Diagnostic Models}
The $Sk_w$ has been linked to many flow statistics, and those links are briefly reviewed because they impose constraints on models quantifying asymmetry in $w'$. They also offer a compact summary of different data sets that enables comparisons across experiments.  

\subsubsection{Cumulant Expansion Models}
The first link to be studied here is the so-called telegraph properties of the $w'$ series and its overall intermittent behaviour.  A third-order cumulant expansion of the individual probability density function (PDF) of the normalized vertical velocity $w_n=w'/\sigma_w$ is introduced and is given by \citep{nakagawa1977prediction,raupach1981conditional,poggi2009flume} 
\begin{align}
\label{eqn:PDF_w}
  {\rm PDF}(w_n)=&{\rm PDF_G} (w_n) \left[1+\frac{1}{6} Sk_w \left(w_n^3-3 w_n \right)\right],\\ {\rm PDF_G}(w_n)=&\frac{1}{\sqrt{2\pi}}\exp\left(-\frac{w_n^2}{2} \right),\nonumber
\end{align}
where ${\rm PDF_G}(.)$ is a zero-mean unit variance Gaussian PDF and the bracketed term corrects the PDF to account for skewness only. This approximation for the PDF enables an estimate of the fraction of time $w'$ is in an updraft ($\Gamma_+$) or downdraft ($\Gamma_-$) and is given by \citep{katul1997turbulent,poggi2009flume,cava2012role,heisel2020velocity}
\begin{align}
\label{eqn:updraft1}
\Gamma_+=\int_{0}^{\infty}{\rm PDF}(w_n) dw_n; ~ \Gamma_-=\int_{-\infty}^{0}{\rm PDF}(w_n) dw_n=1-\Gamma_+, 
\end{align}
which can be integrated using the expansion in equation \ref{eqn:PDF_w} and arranged to yield
\begin{equation}
\label{eqn:updraft}
   Sk_w=\sqrt{72 {\pi}}~ \left(0.5-\Gamma_+\right).
\end{equation}
Here, updrafts and downdrafts are associated with $w'>0$ and $w'<0$, respectively, and no assumptions are made for what concerns the stresses or heat fluxes. An $Sk_w>0$ requires a $\Gamma_+<0.5$ or $\Gamma_{-}>0.5$.  Equation \ref{eqn:updraft} is not prognostic or predictive but suggests that the mean duration of updrafts (or downdrafts) can be explained by asymmetry in the flow.   As shall be seen later, links between the so-called large-eddy skewed turbulence advection velocity used in many CBL parameterizations and $\Gamma_+$ can also be established for shear-only driven flows.

Instead of updraft and downdraft, another measure that characterizes the relative importance of ejections and sweeps (i.e. $\Delta S_o$) on momentum fluxes and linked to $Sk_w$  can be developed using Cumulant Expansion Models (CEM) but on the joint PDF (JPDF) of $w'$ and $u'$.  This measure is defined using quadrant analysis and conditional sampling and is given by \citep{nakagawa1977prediction,raupach1981conditional}
\begin{equation}
\label{dSo}
\Delta S_o = \frac {\langle u'w'\rangle\vert_4 - \langle u'w' \rangle \vert_2}{\overline{u'w'}},
\end{equation}
\noindent where $\langle u'w' \rangle \vert_i$ is a conditional average of events in quadrant $i$, with quadrant 2 corresponding to ejections $(u'<0, w'>0)$ and quadrant 4 to sweeps $(u'>0, w'<0)$.  The statistic $\Delta S_o$ is the fractional difference between contributions of ejection and sweep events to the overall time-averaged flux $\overline{u'w'}$, where the sign indicates whether sweeps ($\Delta S_o>0$) or ejections ($\Delta S_o<0$) are dominant. As before, a third-order cumulant expansion of the JPDF($u',w'$) to link $\Delta S_o$ with key statistical moments can be developed and is given by \citep{raupach1981conditional}
\begin{equation}
\label{dSo_original}
\Delta S_o = \frac{M_{11}+1}{M_{11} \sqrt{2 \pi}} \biggr[ \frac{2
C_1}{(1+ M_{11})^2} + \frac{C_2}{1 + M_{11}}\biggr],
\end{equation}
\noindent where $C_1$ and $C_2$ are defined as
\begin{eqnarray}
C_1 & = & \biggr (1+ M_{11} \biggr) \biggr[\frac{1}{6}(M_{03} - M_{30}) + \frac{1}{2} (M_{21} - M_{12}) \biggr] \nonumber\\
C_2 & = & -\biggr[\frac{1}{6} (2- M_{11} )(M_{03} - M_{30}) +
\frac{1}{2} (M_{21} - M_{12}) \biggr],
\label{Constants_Raupach}
\end{eqnarray}
\noindent and the so-called 'M' notation (i.e. $M_{ij}$) is used to describe different statistical (mixed) moments of $u'$ and $w'$ as
\begin{equation}
\label{M_ij}
M_{ij} = \frac{ \overline {{u'}^i {w'}^j} }{\sigma_u^i
\sigma_w^j}.
\end{equation}
\noindent In M notation, $M_{11}$ defines the correlation coefficient $\overline{w'u'}/(\sigma_u \sigma_w)$, $M_{30}=Sk_u$ and $M_{03}=Sk_w$ define individual skewness values for $u'$ and $w'$, respectively, and $M_{12}$ (associated with wall-normal turbulent transport of flux) and $M_{21}$ (associated with wall-normal turbulent transport of longitudinal velocity variance) define third-order mixed moments. When equation \ref{Constants_Raupach} is inserted into equation \ref{dSo_original}, the final form linking $\Delta S_o$ to the statistical moments is
\begin{equation}
\label{dSo_CEM}
\Delta S_o = \frac{ 1 }{ M_{11} 2 \sqrt{2\pi}} \left[ \frac{M_{11}}{3} \left( M_{03}-M_{30} \right) + \left( M_{21}-M_{12} \right) \right].
\end{equation}
A large corpus of experiments on momentum transport over smooth surfaces and differing types of roughness elements suggest a linear relation between each of the third-order moments. Specifically, $M_{30}=b_u M_{12}$, $M_{03}=b_w M_{12}$, and $M_{21}=b_{uw} M_{12}$ where the respective constant values $b_u \approx 2$, $b_w \approx -1.16$, and $b_{uw} \approx -1$ were presented elsewhere \citep{raupach1981conditional,heisel2020velocity}. The value $b_{uw} \approx -0.6$ was also reported for flows within and just above dense canopies across a wide range of thermal stratification conditions \citep{cava2006buoyancy,katul2018ejective,katul1997turbulent}.  Inserting these linear relations into equation \ref{dSo_CEM} and simplifying yields \citep{heisel2020velocity}
\begin{equation}
\label{structural_model1norm}
Sk_w=M_{03} \approx \frac{{2b_w  \sqrt{2 \pi}}}{ \frac{1}{3}M_{11}(b_w-b_u)+( b_{uw}-1)}M_{11} \Delta S_o.
\end{equation}
In dimensional form, the sought link between $\overline{w'w'w'}$ and $\sigma_w^3 \Delta So$ is now given by
\begin{equation}
\label{structural_model1}
\overline{w'w'w'} \approx \frac{{2 b_w  \sqrt{2 \pi}}}{\frac{1}{3}M_{11}(b_w-b_u)+( b_{uw}-1)}  \sigma_w^3 M_{11} \Delta S_o.
\end{equation}
This relation reveals a $\overline{w'w'w'}$ that is not proportional to $\partial \sigma_w^2/\partial z$ as modeled by gradient-diffusion arguments \citep{launder1975progress} such as the one in equation \ref{eqn:www_gradww} in the absence of large scale adjustment.  Equating this outcome to equation \ref{eqn:updraft} yields an interesting connection between $\Gamma_+$ and $\Delta S_o$ given by 
\begin{align}
\label{structural_model1norm2}
\frac{{2b_w }}{ \frac{1}{3}M_{11}(b_w-b_u)+( b_{uw}-1)}=6 ~C_{CEM}; \\ C_{CEM} M_{11} \Delta S_o=0.5-\Gamma_+. \nonumber
\end{align}
Here, $C_{CEM}$ is a coefficient that is not constant and varies with $z$ due to the z-variations in $M_{11}$.  The latter result is suggestive of a connection between the time fraction in updrafts and downdrafts and the stress fractional imbalance between ejections and sweeps.  When $Sk_w>0$, $0.5-\Gamma_+>0$, ejections occur less than 50$\%$ of the time yet equation \ref {structural_model1norm2} adds an extra finding that ejections dominate momentum transport over sweeps (i.e. $\Delta S_o<0$).

\subsubsection{Realizability Constraints}
One class of models propose that deviations from Gaussian distribution must impact $Sk_w$ as well as the flatness factor $FF_w=\overline{w'^4}/\sigma_w^4$ with some coordination. The 'essence' of this argument is that the mechanisms that generate asymmetry (e.g. $Sk_w$) are not entirely independent of the mechanisms that produce large-scale intermittency (e.g. $FF_w$).  Realizability constraints have been used to guide the development of such coordination between $Sk_w$ and $FF_w$, especially in the CBL. In more detail, two random variables $a'=w'$ and $b'=w'w'$ must satisfy the statistical constraint
\begin{equation}
\label{eq:realize1}
\overline{a'b'}\le \sigma_a \sigma_b, 
\end{equation} 
where $\sigma_a=\sigma_w$ and $\sigma_b=\sigma_{w^2}=\left[\overline{(b'-\overline{b'})^{2}}\right]^{1/2}=\left[\overline{(w'^2-\overline{w'^2})^2}\right]^{1/2}$. Expanding this expression yields
\begin{align}
\label{eq:realize2}
\sigma_{w^2} & =  \left(\overline{w'^4}-(\overline{w'^2})^2\right)^{1/2}=\left[FF_w \sigma_w^4 - (\sigma_w^2)^2\right]^{1/2} \\ & =\sigma_w^2 \left(FF_w-1\right)^{1/2} \nonumber. 
\end{align} 
Inserting this finding into equation \ref{eq:realize1} and re-arranging results in
\begin{equation}
\label{eq:realize3}
{\overline{w'w'w'}}\le \sigma_w^3 \left(FF_w-1\right)^{1/2}.
\end{equation}
Upon dividing both sides by $\sigma_w^3$ and squaring yields $Sk_w^2\le FF_w-1$.
When these inequality constraints are written as equalities with unknown coefficient, they can be expressed as \citep{andre1976turbulence,Alberghi2002}
\begin{equation}
\label{eqn:Skw_FFw}
     FF_w=\alpha_1 (Sk_w^2+1),
\end{equation}
where $\alpha_1$ is a model parameter that should exceed unity to ensure realizability. For a Gaussian PDF, the $Sk_w=0$, $FF_w=3$, and $\alpha_1=3$ ($>1$).  Empirical values ranging from $2.6-3.3$ have been reported across a number of field experiments and LES  \citep{Alberghi2002,maurizi2006dependence}.  This class of models will also be considered using the data here - as it facilitates a new closure scheme for the budgets of $w'^3$ to be derived later on in the prognostic models section.   

\subsubsection{The Fractional Area/Mass Flux Approach}
The $Sk_w$ can also be used to define a so-called turbulent advection velocity $w_a$ using
\begin{equation}
\label{eqn:Skw_def}
    w_a=\frac{\overline{w'^3}}{\sigma_w^2}=Sk_w \sigma_w,
\end{equation}
and this advection velocity is employed to parameterize top-down and bottom-up diffusion in CBLs \citep{zilitinkevich1999third}.  In the eddy advection velocity model, the fractional area occupied by updrafts ($w_+$) and downdrafts ($w_-$) are $a$ and $1-a$.  These fractional areas are presumed constant, which may be plausible in the CBL but not necessarily in the near-neutral ABL (to be explored here). Nevertheless, it is instructive to ask what such a model predicts for $Sk_w$.  For a constant $a$ \citep{abdella2000third}, the conservation of fluid mass leads to the following expressions for the large eddy skewed advection velocity approach: 
\begin{align}
\label{eq:LESTA}
a w_+ + (1-a) w_- = & 0,\\
a w_+^2+ (1-a) w_-^2 = & \sigma_w^2,\\
a w_+^3 + (1-a) w_-^3 = & (Sk_w) \sigma_w^3,\\
a w_+^4 + (1-a) w_-^4 = & (FF_w) \sigma_w^4.
\end{align}
These expressions describe the first four moments and can be solved to yield
\begin{align}
\label{eq:LESTA2}
w_+ = & - \frac{1-a}{a} w_-, ~ {\rm and} ~ w_- =  \pm \frac{\sigma_w}{\sqrt{(1/a)-1}},\nonumber\\
a = & \frac{1}{2} \pm \frac{Sk_w}{2 \sqrt{4 + Sk_w^2}}, ~ {\rm or} ~ Sk_w^2 =-4 +\frac{1}{a}+\frac{1}{1-a},\\
FF_w = & -3 + \frac{1}{a(1-a)}, ~{\rm or} ~ FF_w=1+Sk_w^2. \nonumber
\end{align}
This latter finding is identical to equation \ref{eqn:Skw_FFw} when setting $\alpha_1=1$, which is the minimum $\alpha_1$ necessary for satisfying the realizability constraint as shown in equation \ref{eq:realize3}.  When equating equation \ref{eqn:updraft} to the outcome of equation \ref{eq:LESTA2}, a relation between $a$ (fractional area contribution by updrafts) and $\Gamma_+$ (time fraction of updrafts) can be established, i.e.
\begin{align}
a = \frac{1}{2} \pm \frac{1}{2}\frac{\sqrt{72 {\pi}} \left(0.5-\Gamma_+\right)}{\sqrt{4 + 72 \pi (0.5-\Gamma_+)^2}}.
\end{align}
While $a$ (a spatial quantity) is not linearly related to $\Gamma_+$, a one-to-one correspondence has been established here.  On a similar note, variants on this approach replace $a$ and $1-a$ with PDF($w'>0$) and PDF($w'<0$) as discussed elsewhere \citep{zilitinkevich1999third}. 

\subsection{Prognostic Models}
These models are based on moment expansion of the NSE and closure schemes that link higher- to lower order moments. The prognostic approach commences with the governing equation for $w$.  The first, second, and third order budgets for the statistics of $w'$ are reviewed here along with conventional closure models employed. Proposed revisions to them, especially for the third order budgets are also presented. For an incompressible or constant density flow, the instantaneous equation for $w$ (or $w'$) in the absence of stratification and Coriolis is given by
\begin{equation}
\label{eq:w_inst}
\frac{\partial w} {\partial t}+u\frac{\partial w} {\partial x}+v \frac{\partial w} {\partial y}+w\frac{\partial w} {\partial z}=-\frac{1}{\overline{\rho}}\frac{\partial P}{\partial z} -g+\nu \frac{\partial^2 w} {\partial x_j \partial x_j},
\end{equation}
where $t$ is time, $\rho$ is the fluid density (constant here), $P$ is the pressure, and repeated indices imply summation.  

\subsubsection{First Moment}
Upon averaging equation \ref{eq:w_inst} and considering stationary (i.e. $\partial \overline{(.)}/{\partial t}=0$) and planar homogeneous (i.e. $\partial \overline{(.)}/{\partial x_1}=\partial \overline{(.)}/{\partial x_2}=0$) flow in the absence of subsidence (i.e. $\overline{w}=0$), the vertical velocity variance is given by  
\begin{equation}
\label{eq:w_avg}
\frac{1}{2}\overbrace{\frac{\partial \overline{w'w'}} {\partial z}}^{\rm Inertial ~ Term}=\underbrace{-\frac{1}{\overline{\rho}}\frac{\partial \overline{P}}{\partial z}}_{\rm Pressure~Gradient} 
\overbrace{-g}^{\rm Gravitational ~ Acceleration},
\end{equation}
which upon integration with respect to $z$ yields a Bernoulli-like equation
\begin{equation}
\label{eq:w_bernoulli}
\frac{1}{2}\overline{w'w'}+\frac{\overline {P}}{\overline{\rho}}+g z=C_B,
\end{equation}
where $C_B$ is a constant of integration.  When $\overline{P}=-\overline{\rho} g z$ (i.e. hydrostatic approximation), $\partial \sigma_w^2/\partial z=0$ or $\sigma_w$ must be a constant independent of $z$.  Thus, deviations from a constant $\sigma_w^2$ in $z$ signifies deviations from hydro-static assumption in the mean pressure vertical gradient.

\subsubsection{Second Moment}
The budget equation for $\overline{w'w'}$ can be similarly derived by multiplying equation \ref{eq:w_inst} by 2$w'$ and averaging to yield
\begin{equation}
\label{eq:wvar_budget}
\overbrace{\frac{\partial \overline{w'w'w'}}{\partial z}}^{\rm Inertial~Term}=\underbrace{-2\overline{w'\frac{\partial p'}{\partial z}}}_{\rm Pressure-Velocity~Interaction}+2 \underbrace{\nu \overline{w' \frac{\partial^2 w'} {\partial x_j \partial x_j}}}_{\rm Viscous~Destruction},
\end{equation}
where $p'$ is turbulent pressure deviation from the mean or hydrostatic state.  This budget identifies two mechanisms that impact $\partial \overline{w'^3}/\partial z$: The first is a pressure-velocity interaction term that can act as a source or a sink as later discussed and the second is a viscous destruction term that acts as a sink term.  A model for the pressure-velocity interaction term is needed to mechanistically link the skewness to turbulent processes.  This term may be modeled using a linear return-to-isotropy Rotta scheme \citep{Rotta1951} modified here to maintain the pressure transport term \citep{canuto1994second}.  It is given by \citep{Rotta1951,bou2018role,canuto1994second}
\begin{equation}
\label{eq:Rotta}
-2\overline{w'\frac{\partial p'}{\partial z}}=\frac{C_R}{\tau}\left(\frac{2K}{3} -\overline{w'^2} \right)+2 \frac{\partial }{\partial z} \overline{w'p'},
\end{equation}
where $\tau$ is a time scale, $C_R=1.8$ is the Rotta constant, $K=(1/2) \overline{u'_i u'_i}$ is the turbulent kinetic energy at $z$.  The Rotta model actually applies to $\overline{p'\partial w'/\partial z}$ (i.e. no gradients in $p'$ emerge).  However, $\partial (\overline{w'p'})/\partial z=\overline{p'\partial w'/\partial z} + \overline{w' \partial p'/\partial z}$, which is why the pressure transport term emerges in equation \ref{eq:Rotta}.  The term $\partial (\overline{w'p'})/\partial z$ has been ignored in many atmospheric turbulence studies \citep{stull2012introduction,wyngaard_2010} but can be included if finite and known \citep{su2023large}.

At large scales, the redistribution of kinetic energy among velocity components by the pressure term to achieve an equi-partition state is reasonably established \citep{hanjalic2021reassessment,Pope2000,Stiperski2021}.  Because the viscous dissipation rate of $K$ ($=\epsilon > 0$) occurs at small (or micro) scales that are presumed to be isotropic, it is reasonable to assume that $\epsilon_u=\epsilon_v=\epsilon_w=(1/3)\epsilon$, where $\epsilon_u$, $\epsilon_v$, $\epsilon_w$ are the viscous dissipation rates of $\overline{u'u'}$, $\overline{v'v'}$, and $\overline{w'w'}$.  Thus, with a return to isotropy closure and an isotropy of dissipation rate of $K$, a model for $\overline{w'^3}$ may be constructed and is given as \citep{bou2018role}
\begin{equation}
\label{eq:wvar_model}
\frac{\partial \overline{w'w'w'}}{\partial z}=\frac{C_R}{\tau}\left(\frac{2 K}{3} -\overline{w'^2} \right)-\frac{1}{3}\epsilon.
\end{equation} 
In the original Rotta scheme, $\tau=2K/\epsilon$ (i.e. return-to-isotropy time scale is proportional to the relaxation time scale via $C_R$) and equation \ref{eq:wvar_model} can be expressed in non-dimensional form as
\begin{equation}
\label{eq:wvar_nondim}
\frac{1}{\epsilon}\frac{\partial \overline{w'w'w'}}{\partial z}={C_R}\left(\frac{1}{3} - \frac{\overline{w'^2}}{2 K} \right)-\frac{1}{3}.
\end{equation} 
It is instructive to inquire about a reference state that result in a zero vertical gradient in $\overline{w'^3}$ similar to the analysis leading to a constant $\sigma_w^2$ with $z$ variations.  From the analysis above, this state is achieved when the anisotropy measure $A_w={\sigma_w^2}/{(2 K)}$ is a constant given by an equilibrium value
\begin{equation}
\label{eq:zerograd}
A_{w,e}=\frac{1}{3}-\frac{1}{3 C_R}.
\end{equation} 
For a $C_R=1.8$ (or a $C_R=0.9$ when $K$ is used in lieu of $2K$), $A_{w,e}=0.15$.  In the inertial sublayer (ISL) of the atmosphere (or near-neutral ASL), common values for the second moments are $\sigma_u/u_*=2.4$, $\sigma_v/u_*=2.1$, and $\sigma_w/u_*=1.25$ so that typical ISL values for $A_w=0.13$, close to the predicted value ($=0.15$) from $C_R$ only.  Thus, it is expected that $\overline{w'^3}$ be a constant independent of $z$ as these idealized conditions are approached.  Equation \ref{eq:wvar_nondim} makes clear that a reduction in $A_w$ below $A_{w,e}$ leads to $\partial \overline{w'^3}/\partial z>0$. In general, wall-blocking in the RSL and contributions from large eddies to $K$ in the outer layer tend to reduce $A_w$ below $A_{w,e}$.   Thus, in both regions, $\partial \overline{w'^3}/\partial z>0$. 

For completeness, deviations from small-scale isotropy (i.e. $\epsilon_u$=$\epsilon_v$=$\epsilon_w$) can be accommodated in this framework as these conditions may occur when the Reynolds numbers are not high.  If small-scale anisotropy does persist \citep{spalart1988direct}, then it is convenient to express $\epsilon_u+\epsilon_v=\alpha_d \epsilon_w$, where $\alpha_d<2$ as discussed elsewhere \citep{bou2018role}.  For such anisotropy, $\epsilon_w=\epsilon/(1+\alpha_d)$ and 
\begin{equation}
\label{eq:wvar_nondim2}
\frac{1}{\epsilon}\frac{\partial \overline{w'w'w'}}{\partial z}={C_R}\left(\frac{1}{3} - \frac{\overline{w'^2}}{q^2} \right)-\frac{1}{1+\alpha_d}.
\end{equation} 
This implies that $A_{w,e}$ must be modified to $A_{w,ani}$ so as to include small-scale anisotropy in the turbulent kinetic energy dissipation rates.  These corrections are not pursued further given the uncertainty in the dissipation rate estimates in the data sets used here. 

\subsubsection{Third Moment}
The budget for the time rate of change of $\overline{w'w'w'}$ can be derived by multiplying equation \ref{eq:w_inst} with $3w'^2$ and averaging the outcome to yield \citep{canuto1994second,zeman1976modeling,buono2024vertical}
\begin{align}
\label{eq:budgetwww1}
\frac{\partial \overline{w'^3}}{\partial t}=0&=\overbrace{-\frac{\partial \overline{w'w'^3}}{\partial z}+3 \overline{w'w'}\frac{\partial \overline{w'w'}}{\partial z}}^{\rm Inertial~Term}\\&\underbrace{\frac{-3}{\overline{\rho}}\left( \overline{w'w' \frac{\partial p'}{\partial z}}\right)}_{\rm Pressure-Velocity~Interaction} \underbrace{-2 \nu \left( 3 \overline{w' \frac{\partial w'}{\partial z} \frac{\partial w'}{\partial z}}\right)}_{\rm Viscous~Destruction}\nonumber.
\end{align} 
To close the fourth moment, a number of possibilities exist.  To maintain generality, the inertial term can be formulated as  
\begin{align}
\label{eq:budgetwww2}
-\frac{\partial \overline{w'^4}}{\partial z}+3 \overline{w'w'}\frac{\partial \overline{w'w'}}{\partial z} & =  -\sigma_w^4\frac{\partial FF_w}{\partial z}\\ &+\sigma_w^2 \left(3-2FF_w\right) \frac{\partial \sigma_w^2}{\partial z}\nonumber.
\end{align} 
The most common closure is the quasi-Gaussian approximation setting $FF_w=3$ without making any assumptions about the asymmetry \citep{andre1976turbulence}. In this case, $\partial FF_w/\partial z=0$ and the inertial term reduces to $-3 \sigma_w^2 (\partial \sigma_w^2/\partial z)$.
Another option is to use the result of equation \ref{eqn:Skw_FFw}.  As simplification, it is assumed that the first term on the right-hand side of equation \ref{eq:budgetwww2} is much smaller than the second term so that:
\begin{align}
\label{eq:budgetwww2s}
-\frac{\partial \overline{w'^4}}{\partial z}+3 {\sigma_w^2}\frac{\partial \sigma_w^2}{\partial z} = \sigma_w^2 \left[3- 2\alpha_1 \left(Sk_w^2+1\right)\right] \frac{\partial \sigma_w^2}{\partial z}.
\end{align} 
Interestingly, when $Sk_w^2<<1$, and $\alpha_1$ does not deviate appreciably from 3, the Gaussian approximation is recovered.  Thus, the closure in equation \ref{eq:budgetwww2s} suggests that some deviations from Gaussian can be accommodated provided $FF_w$ does not vary appreciably with $z$.

Closure models for the pressure-velocity interaction and viscous dissipation terms have been proposed \citep{zeman1976modeling,canuto1994second,buono2024vertical}. A linear return to isotropy scheme yields \citep{Rotta1951}
\begin{equation}
-\frac{3}{\overline{\rho}}\left( \overline{w'w' \frac{\partial p'}{\partial z}}\right)=\frac{3}{2}\frac{C_R}{\tau'}\left(\frac{\overline{w'q'}}{3} -\overline{w'w'^2} \right),
\end{equation}
and the viscous dissipation contribution can be modeled as
\begin{equation}
2 \nu \left( 3 \overline{w' \frac{\partial w'}{\partial z} \frac{\partial w'}{\partial z}}\right)=2 \overline {\epsilon' w'}= c_2 \frac{\overline{w'q'}}{\tau'},
\end{equation}
where $c_2$ is a similarity constant, $q'=u_i'u_i'$, $\epsilon'$ is the fluctuating dissipation rate, and $\tau'$ is a decorrelation time scale that need not be identical to $\tau$ because $w'$ and the instantaneous time scale $\epsilon'/q'$ can be correlated.  Here, the interaction between $w'$ and the pressure transport term has been ignored though its effect can be accommodated if necessary.

Inserting these approximations into equation \ref{eq:budgetwww1} yields \citep{buono2024vertical}, 
\begin{equation}
\label{eq:budgetwww_cl1}
\overline{w'^3}=\overbrace{-\frac{2\tau' \sigma_w^2}{3 C_R} \frac{\partial \overline{w'w'}}{\partial z}}^{\rm Gradient-Diffusion}+\overbrace{\overline{w'q'} \left( \frac{1}{3} - \frac{2 c_2}{3 C_R}\right)}^{\rm Non-local~Transport}.
\end{equation} 
For operational purposes, a model for $\overline{w'q'}$ is needed to prognostically determine $Sk_w$.  A plausible closure that has been extensively studied is \citep{lopez1999wall,buono2024vertical}
\begin{equation}
\label{eq:budgetwww3}
\overline{w'q'}=- \kappa z u_* \phi_L(z) \frac{\partial (2 K)}{\partial z},
\end{equation} 
where $\kappa=0.4$ is the von Ka\'rma\'n constant, and $\phi_L$ is a correction to accommodate outer layer effects \citep{blackadar1962vertical}.  It is common in open channel flows to assume $\phi_L(z)=1-z/\delta$. In atmospheric boundary layers and wind tunnels, $\phi_L(z)=(0.5+{\kappa z}/{\alpha_m \delta})^{-1}$, where $\alpha_m=\kappa/4$.  The quadratic variation in $z$ of the eddy diffusivity has been proposed in earlier modeling studies of momentum transfer \citep{o1970note} and tested for open channel flows \citep{lopez1999wall}.  For $z/\delta<<0.1$, the eddy diffusivity increases linearly with $z$.  However, the quadratic term in $z$ becomes the dominant term as $z/\delta$ increases beyond 0.5.  For wind tunnels, $\phi_L$ increases monotonically and approaches a constant (and maximum) value. Inserting this closure into equation \ref{eq:budgetwww_cl1} yields \citep{buono2024vertical}
\begin{equation}
\label{eq:budgetwww4}
Sk_w=-\frac{2}{3 C_R}\left[A_{t,w} \frac{\partial \overline{w'w'}}{\partial z}+B_{t,u}\frac{\partial K}{\partial z}\right], 
\end{equation} 
where 
\begin{equation}
\label{eq:Kw_Ku}
A_{t,w}= \frac{\tau'}{\sigma_w }; B_{t,u}=\left(C_R- {6 c_2} \right)\frac{\kappa z u_*\phi_L(z)}{\sigma_w^3}.
\end{equation} 
Because $K=(1/2) (\sigma_u^2 + \sigma_v^2 + \sigma_w^2)$ and $\sigma_u^2 \approx \sigma_v^2 + \sigma_w^2$, it follows that $K \approx \sigma_u^2$ and $\partial K/\partial z$ can be replaced by $\partial \sigma_u^2/\partial z$. It is worth noting that, even within a near-neutral ASL, $\sigma_u$ is impacted by eddies much larger than $z$ consistently with  Townsend's attached eddy hypothesis \citep{banerjee2013logarithmic,marusic2013logarithmic,townsend1980structure,buono2024vertical}. 
A plausible choice is $\tau' = \kappa z u_* \phi_L(z)/u_*^2$, which makes the two eddy viscosity formulations for $A_{t,w}$ and $B_{t,u}$ comparable in magnitude in the ISL and outer layer, as routinely done in turbulence closure schemes.

\section{Data Sets}
The data sets have been collected over smooth walls \cite{manes2011turbulent,peruzzi2020scaling,poggi2002experimental}, rough walls \cite{heisel2020velocity}, and permeable walls with varying permeability and Reynolds numbers \cite{manes2011turbulent}.  Table \ref{tab:table1} summarizes the experimental conditions and the data sets as well as the original sources describing the experiments. Data sets came from both open channel flows (OC) and wind tunnel (WT) experiments. Longitudinal and wall normal flow velocities have been measured with laser doppler anemometry in OC and with cross-hotwire anemometry in WT experiments. Permeable wall experiments have been performed with bed porosity measured in pores per inches (ppi) of 60 (MNP1, MNP2), 30 (MNP3, MNP4) and 10 (MNP5). Rough wall experiments (HLR1, HLR2) have been performed with a woven wire mesh with a roughness length of 6 mm. More methodological details are provided in the original sources presented in Table \ref{tab:table1}.  These data sets represent canonical wall-bounded flows and lack certain complexities that are present in the ABL even under neutral conditions and in the absence of Coriolis effects. The chief complexity absent here is the effect of a stratified capping inversion that influences the outer region behavior of the conventionally neutral ABL. This effect is likely to modify, at minimum, the $\phi_L(z)$ formulation. 

\begin{table*}
\caption{Summary of the data sets used in the analysis. Bed types are classified as {S} smooth, {P} porous, and {R} rough. Flow types are reported as {OC} for open channels, and {WT} for wind tunnels. For convenience, the symbols used in the figures are reported here as well.}

\begin{ruledtabular}
\begin{tabular}{cccccccc}
        Source & Data set & Bed & Flow & $\delta \times 10^{-3} [m]$ & $u_* \times 10^{-3} [m/s]$ & $Re_{\tau}$ & Symbol\\ \hline
        \citet{manes2011turbulent}    & MNS  & S & OC &  60 &  41 &  2160 & \multirow{9}{*}{\includegraphics[scale=0.303]{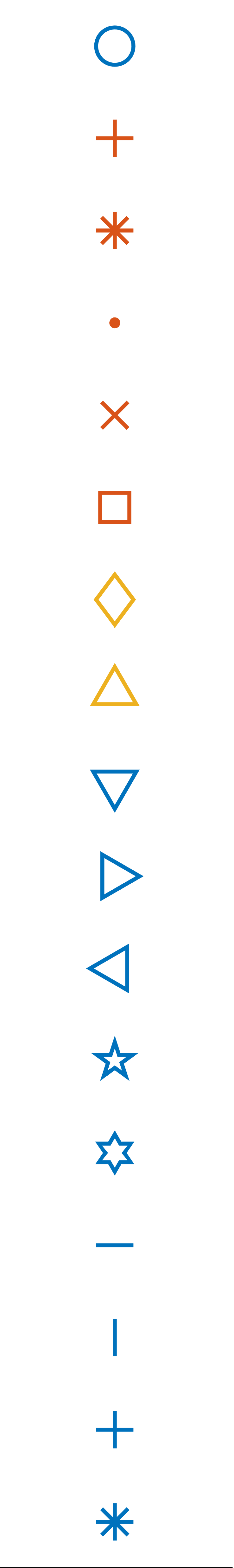}} \\ 
                                    ~ & MNP1 & P & OC &  96 &  28 &  2349 & \\  
                                    ~ & MNP2 & P & OC & 110 &  34 &  3234 & \\  
                                    ~ & MNP3 & P & OC & 115 &  18 &  1856 & \\  
                                    ~ & MNP4 & P & OC & 146 &  46 &  5840 & \\  
                                    ~ & MNP5 & P & OC &  89 &  49 &  3848 & \\ \hline 
        \citet{heisel2020velocity}    & HLR1 & R & WT & 408 & 370 &  9611 & \\ 
                                    ~ & HLR2 & R & WT & 391 & 550 & 13683 & \\
                                    ~ & HLS1 & S & WT & 222 & 260 &  3681 & \\
                                    ~ & HLS2 & S & WT & 203 & 350 &  4536 & \\ \hline  
        \citet{peruzzi2020scaling}    & PRS1 & S & OC & 200 &  10 &  1730 & \\ 
                                    ~ & PRS2 & S & OC & 120 &   8 &   795 & \\ 
                                    ~ & PRS3 & S & OC &  85 &  22 &  1657 & \\ \hline
        \citet{poggi2002experimental}        & PGS1 & S & OC &  50 &  21 &  1071 & \\ 
                                    ~ & PGS2 & S & OC &  45 &   7 &   331 & \\ 
                                    ~ & PGS3 & S & OC &  42 &  30 &  1232 & \\ 
                                    ~ & PGS4 & S & OC &  46 &  19 &   845 & \\ 
        
    \end{tabular}
    \end{ruledtabular}
\label{tab:table1}
\end{table*}

\section{Results}
\begin{figure}
\centering
\noindent\includegraphics[width=20pc]{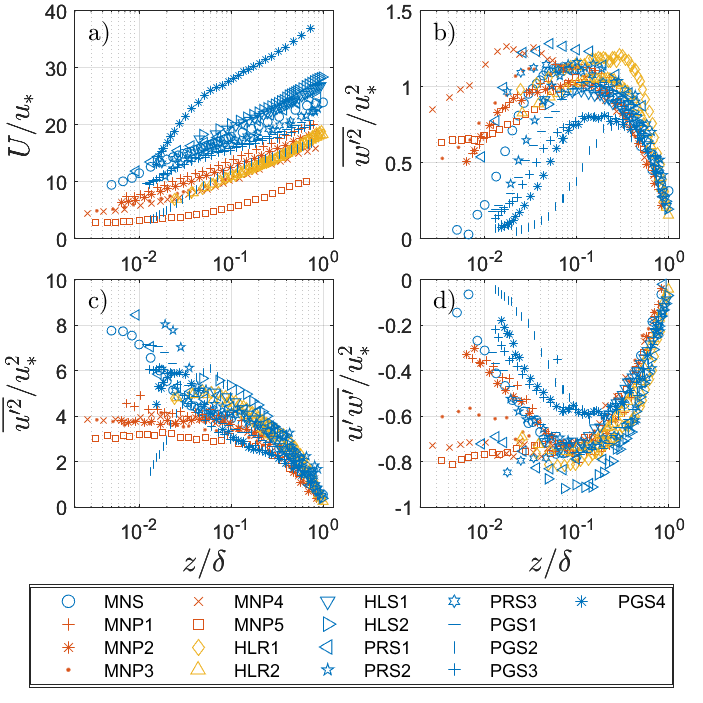}
\caption{Measured values of: (a) normalized mean velocity $U/u_*$, (b) vertical velocity variance $\overline{w'^2}/u_*^2$, (c) longitudinal velocity variance $\overline{u'^2}/u_*^2$, and (d) turbulent stress $\overline{u'w'}/u_*^2$ as a function of the normalized wall normal distance $z/\delta$. In this and all later figures, the symbols correspond to the data sets in Table \ref{tab:table1}.}
\label{Fig:MeanFlow}     
\end{figure}

The results section investigates the diagnostic models first and then proceeds to explore the prognostic model predictions using the third order budget.  Before doing so, the measured first and second moments are reported and discussed in Figure \ref{Fig:MeanFlow} for completeness. 
The normalized mean velocity as a function of $z/\delta$ (i.e. presented with outer layer variables) shows large differences across experiments due to simultaneous Reynolds number and roughness effects. No expected collapse of the data onto a single curve is expected when using such outer scaling variables for $U/u_*$ (Figure \ref{Fig:MeanFlow}a).   Much of the scatter in the second moments is deemed to be below the inertial sublayer (ISL) - roughly below $z/\delta=0.08$ (Figure \ref{Fig:MeanFlow}b-d).  A region of constant stress and constant $\sigma_w^2$ exists for $z/\delta\in[0.08,0.2]$ across most of the data sets (Figure \ref{Fig:MeanFlow}b).  This region delineates operationally the ISL.  For $z/\delta>$0.3, the $\sigma_w^2/u_*^2$, $\sigma_u^2/u_*^2$, and $\overline{u'w'}/u_*^2$ decline in magnitude with increasing $z/\delta$, which is of significance to prognostic models explaining the sign of $Sk_w$ and the dominant processes controlling its magnitude.  

\subsection{Cumulant Expansion Models}
\begin{figure}
\centering
\noindent\includegraphics[width=20pc]{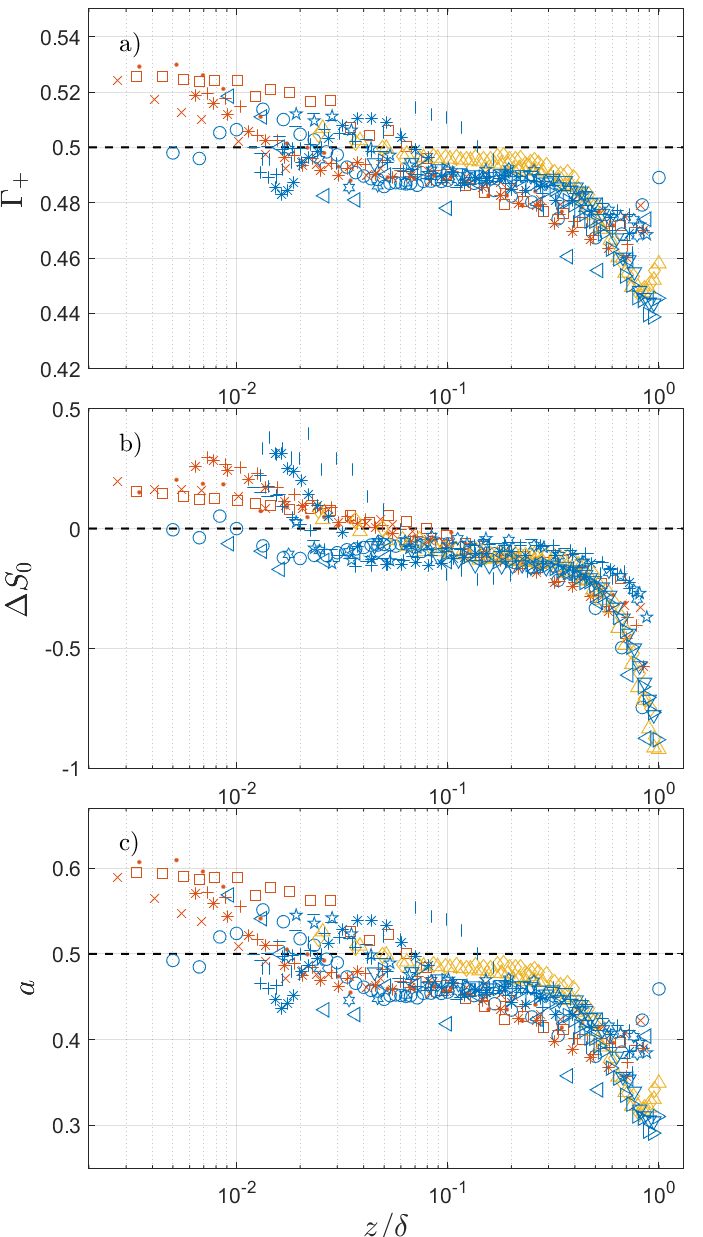}
\caption{Measured values of: (a) the fraction of time $w'>0$ (or $\Gamma_+$), (b) the imbalance between sweeps and ejection contributions to the Reynolds stress (or $\Delta S_o$), and (c) the fractional area $a$ of updrafts as a function of the normalized wall normal distance $z/\delta$. The horizontal dashed lines represent the conditions whereby updrafts and downdrafts are perfectly balanced.}
\label{Fig:dSoGamma}     
\end{figure}

Figure \ref{Fig:dSoGamma} presents the measured values of $\Gamma_+$, $\Delta S_o$, and $a$ as a function of $z/\delta$.  
For $z/\delta>0.08$, all data sets suggest $0.5-\Gamma_+ >0$.  In fact, for $z/\delta\in[0.1,0.3]$,$0.48< \Gamma_+ < 0.50$ for all datasets except one, which is approximately constant with respect to $z$ variations.  The collapse of the data sets for $\Delta S_o$ is also rather remarkable when inspecting the same interval $z/\delta\in[0.1,0.3]$.  For $z/\delta > 0.07$, all data sets suggest ejections dominate ($\Delta S_o <0$) consistent with rough-wall wind tunnel \citep{raupach1981conditional} and smooth-wall open channel flow  \citep{nakagawa1977prediction} experiments.  As $z/\delta >0.9$, $\Delta S_o$ becomes ill-defined because the turbulent stress is small.  Last, it is noted that the predicted $a$ appears independent of $z$ for $z/\delta\in[0.1,0.3]$, while it is substantially reduced when $z/\delta>0.3$ for the majority of datasets. However, the most consistent behaviour in terms of 'data collapse' identifying the strength of ejections across datasets appears to be $\Delta S_o$ (as expected).
\begin{figure}
\centering
\noindent\includegraphics[width=20pc]{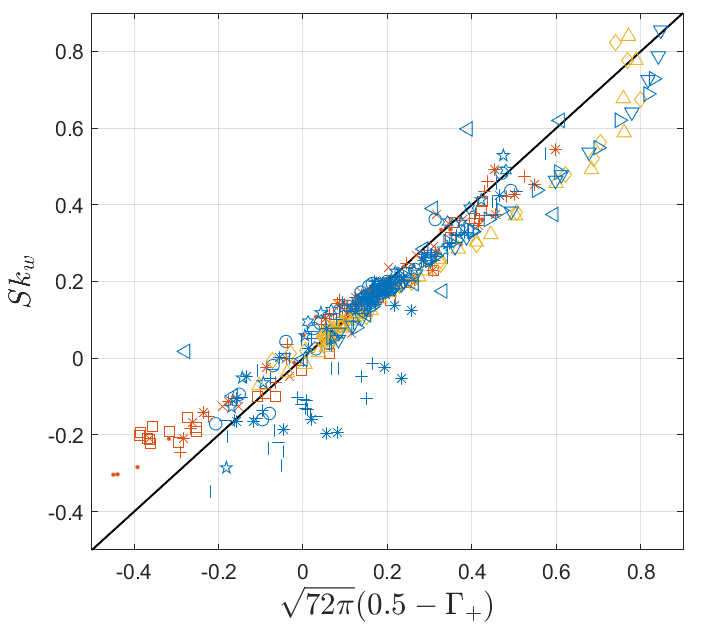}
\caption{Comparison between vertical velocity skewness ($Sk_w$) modeled using third-order cumulant expansion (via measured  $\Gamma_+$) and measured $Sk_w$. All values of $z/\delta$ are included.}
\label{Fig:Skw_Gamma}     
\end{figure}

\begin{figure}
\centering
\noindent\includegraphics[width=20pc]{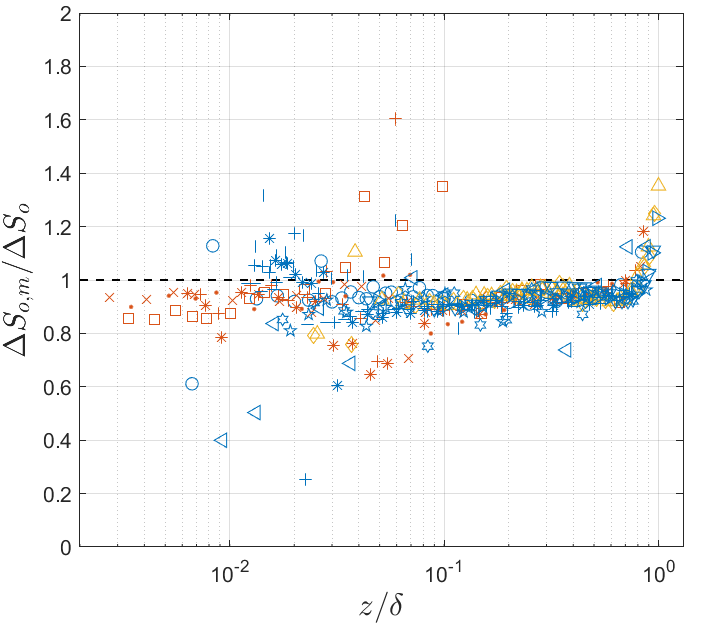}
\caption{Ratio of $\Delta S_{o,m}$ modeled using the third-order cumulant expansion and measured $\Delta S_o$ (computed from quadrant analysis) as a function of $z/\delta$.}
\label{Fig:dSo_z_delta}     
\end{figure}

\begin{figure}
\centering
\noindent\includegraphics[width=20pc]{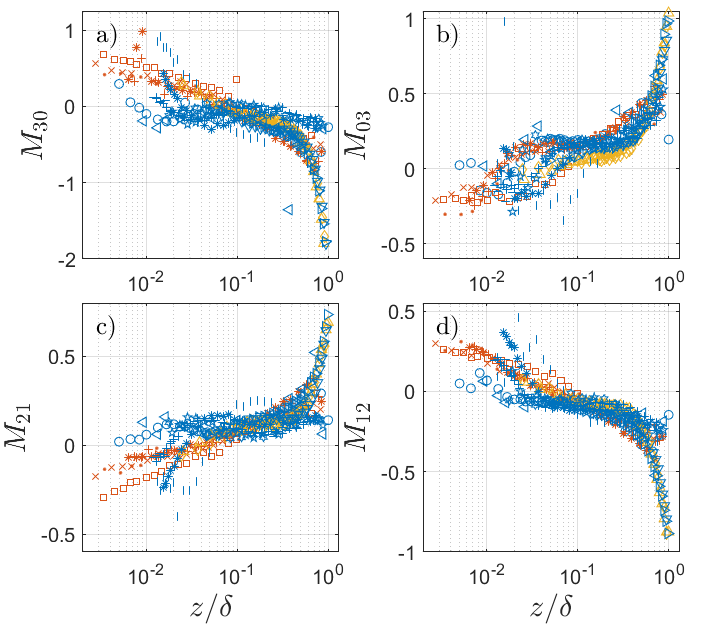}
\caption{The variations of the triple moments $M_{ij}$ as a function of the normalized wall normal distance $z/\delta$.}
\label{Fig:Moments_CEM}     
\end{figure}

\begin{figure}
\centering
\noindent\includegraphics[width=20pc]{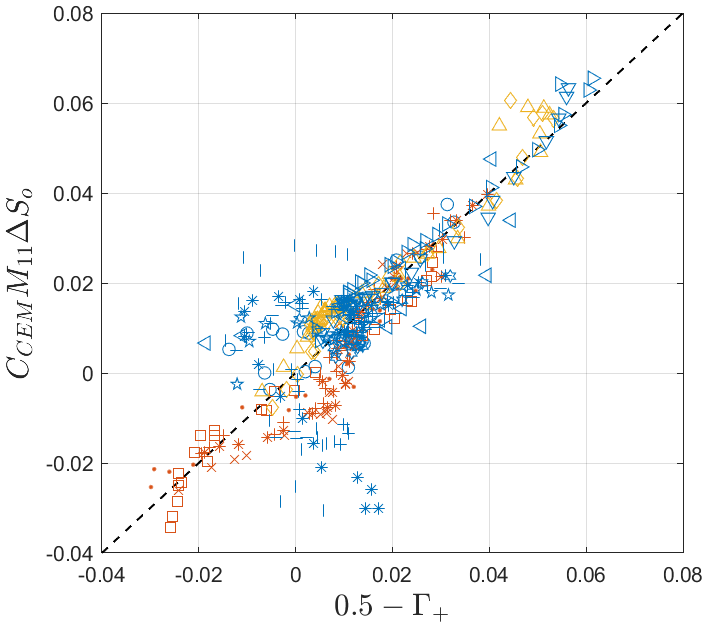}
\caption{Comparison between $0.5-\Gamma_+$ and measured $C_{CEM} M_{11} \Delta S_o$.  The one-to-one line is based on the simplified CEM representation used to obtain equation \ref{structural_model1norm2}.  }
\label{Fig:M11dSo_Gamma}     
\end{figure}

The relation between measured $Sk_w$ and predictions from $\Gamma_+$ using equation \ref{eqn:updraft} (i.e. using CEM) is shown in Figure \ref{Fig:Skw_Gamma}. The agreement is acceptable considering that this comparison covers all sublayers including the buffer- and roughness- sublayers.  As predicted by equation \ref{eqn:updraft}, when $0.5-\Gamma_+>0$, the $Sk_w>0$, and conversely. Likewise, the ratio of $\Delta S_o$ modeled using a third order CEM applied to the JPDF($w'$,$u'$) to measured $\Delta S_o$ as a function of $z/\delta$ is also shown in Figure \ref{Fig:dSo_z_delta}. Once again, the agreement is acceptable for much of the boundary layer region ($z/\delta\in[0.1,0.9]$).  
The values of the individual moments $M_{ij}$ used in the determination of $\Delta S_{o,m}$ (modeled using CEM) as a function of $z/\delta$ are presented in Figure \ref{Fig:Moments_CEM}.
The most consistent collapse across all data sets is for $M_{03}=Sk_w$ in the region of $0.1 <z/\delta<0.9$ (Figure \ref{Fig:Moments_CEM}b) followed by $M_{12}$ (Figure \ref{Fig:Moments_CEM}d). Moreover, there is a notable collapse of $M_{03}=Sk_w$ to a near-constant positive value for $z/\delta\in[0.08, 0.3]$, namely in the ISL, consistent with previous empirical studies \citep{chiba1978stability}. The height independence of $M_{12}$ was proposed to delineate the ISL in some studies on rough-wall turbulence \citep{lopez1999wall} though it appears from the analysis here that $M_{03}$ may be an acceptable single-variable substitute.  Across all data sets and regions, the $C_{CEM} M_{11} \Delta S_o$ appears to vary linearly with $0.5-\Gamma_+$ as shown in Figure \ref{Fig:M11dSo_Gamma}.  This linearity is consistent with predictions from equation \ref{dSo_CEM} and suggests that $Sk_w$, $\Gamma_+$, and $C_{CEM} M_{11} \Delta S_o$ are linearly related as predicted from third order CEMs (see equation \ref{structural_model1norm2}). A main finding is that $\Gamma_+$ contains significant information about imbalances between sweeps and ejections responsible for momentum transfer.

\subsection{Realizability Constraints}
\begin{figure}
\centering
\noindent\includegraphics[width=20pc]{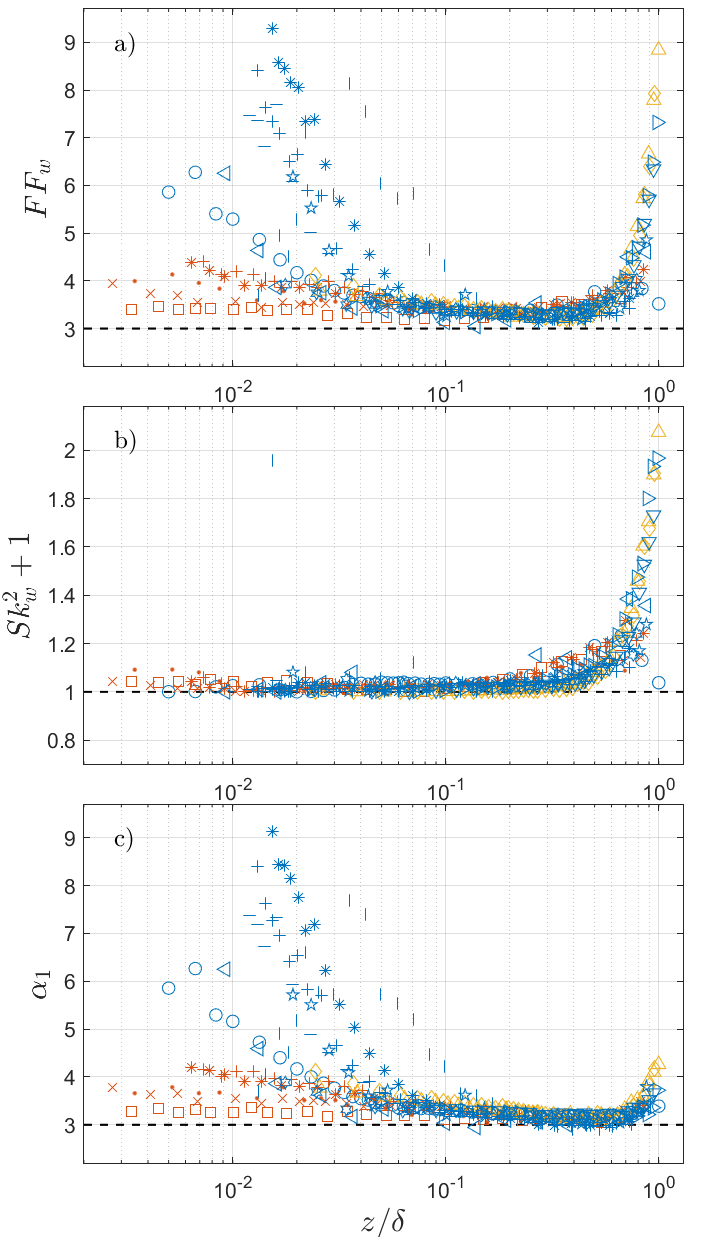}
\caption{The variations of: (a) measured $FF_w$, (b) measured $Sk_w^2+1$, and (c) $\alpha_1$ estimated from $FF_w$ and $Sk_w^2+1$ as a function of the normalized wall normal distance $z/\delta$. The horizontal dashed lines represent the Gaussian approximation.}
\label{Fig:Realizability}     
\end{figure}

In general, the skewness and flatness factors of a PDF are independent quantities.  However, in turbulence modeling the nature of the second-order non-linearity of the Navier-Stokes equation  means that budgets for the statistical moment $m$ of a single flow variable such as $w'$, require moment $m+1$ to be known.  Therefore, it may be conjectured that the physics of turbulence requires some coordination between $FF_w$ and $Sk_w$.  The realizability constraint as formulated here links $Sk_w$ to $FF_w$ by replacing inequalities with equalities along with associated coefficients such as $\alpha_1$. It is important to note that this inequality constraint is only statistical (i.e. applies to any random variable) and replacing inequalities with equalities is not derived from the physics of turbulence. Nonetheless, Figure \ref{Fig:Realizability}c shows the variations in $\alpha_1$ using measured $Sk_w$ and $FF_w$ as a function of $z/\delta$ (reported in Figure \ref{Fig:Realizability}a-b).  For $0.07 <z/\delta<0.9$, the values of $\alpha_1$ are around 3.3 for all data sets.  This finding supports the working assumption that the inertial term in the third-order budget of $\overline{w'^3}$ reduce to $(3-2 \alpha_1)\sigma_w^2 (\partial \sigma_w^2/\partial z)$. Moreover, an $\alpha_1=3.3$ is sufficiently close to a Gaussian value when modeling the entire inertial term using a Gaussian approximation (i.e. $\alpha_1=3$) and deviates appreciably from the prediction by the fractional area/mass flux approach that leads to $\alpha_1=1$.  

\subsection{Gradient-Diffusion Prognostic Models}
\begin{figure}
\centering
\noindent\includegraphics[width=20pc]{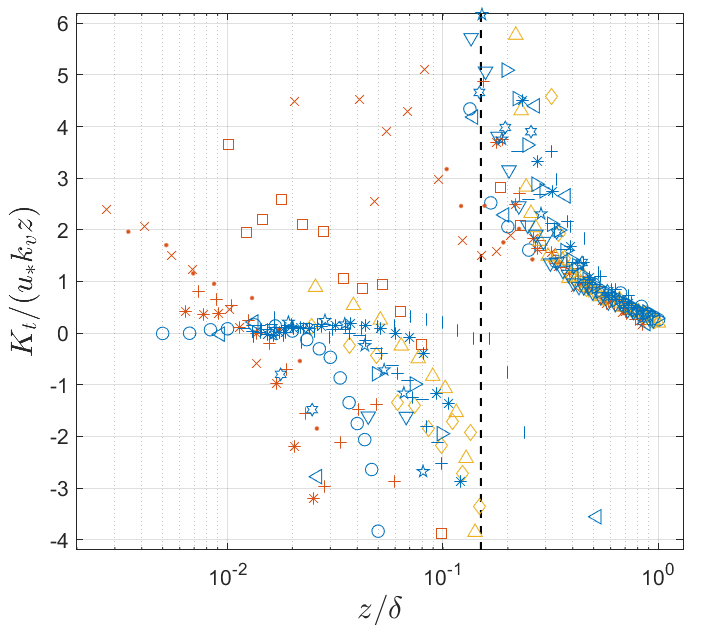}
\caption{The variation of the normalized computed eddy diffusivity $K_t$ as a function of the normalized wall normal distance $z/\delta$. The vertical dashed line roughly indicates $\partial \sigma_w^2/\partial z=0$.}
\label{Fig:grad_diff}     
\end{figure}

\begin{figure}
\centering
\noindent\includegraphics[width=20pc]{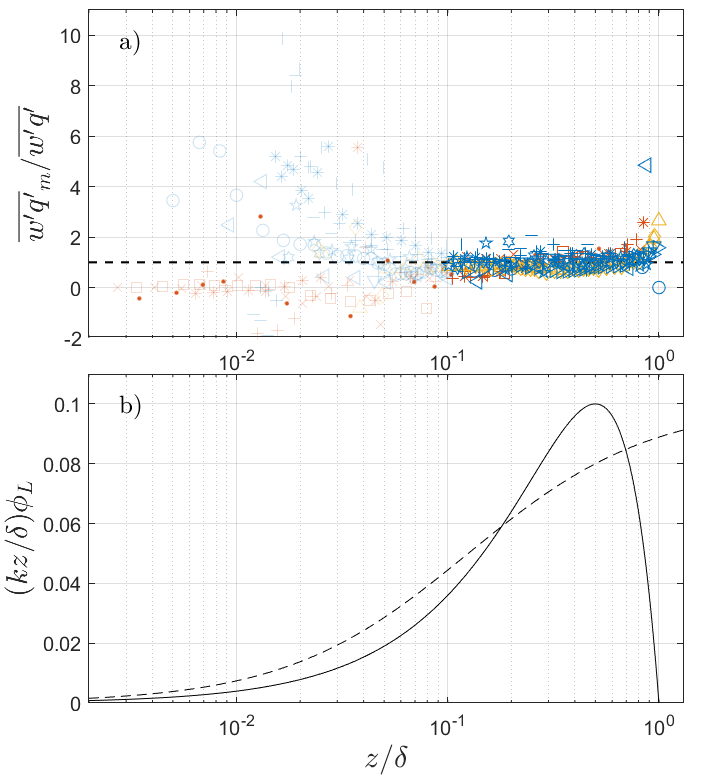}
\caption{(a) Comparison between measured and down-gradient modeled vertical transport of turbulent kinetic energy when setting $q_m=u'^2$. The shaded markers indicate $z/\delta<0.1$. The horizontal dashed line denotes a perfect agreement between model and data. (b)  The assumed effective dimensionless mixing length $\kappa (z/\delta) \phi_L(z/\delta)$ for open channel flows (solid line) and wind tunnels (dashed line). Note the parabolic behavior for open channel flows (characterised by a free surface) and the monotonically increasing values for wind tunnel studies.}
\label{Fig:wq_gradK}     
\end{figure}

\begin{figure}
\centering
\noindent\includegraphics[width=20pc]{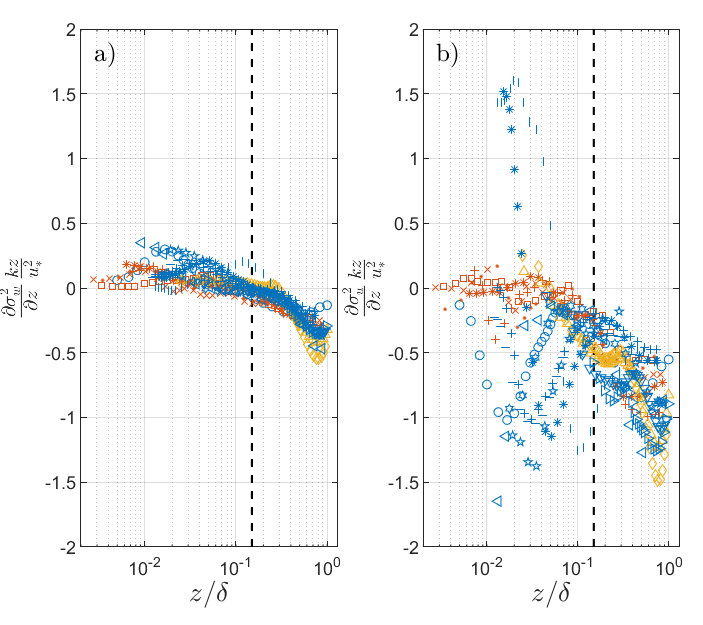}
\caption{Normalized (a) vertical and (b) horizontal velocity gradients as a function of the normalized wall normal distance $z/\delta$. The vertical dashed lines roughly indicate the upper limit of the ISL.}
\label{Fig:gradComp}     
\end{figure}

\begin{figure}
\centering
\noindent\includegraphics[width=20pc]{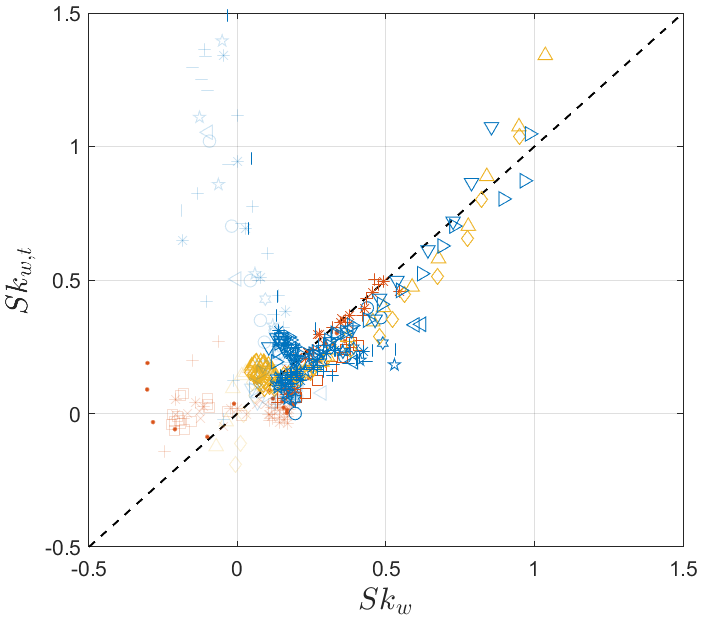}
\caption{Comparison between measured $Sk_w$ and predictions from equation \ref{eq:Skw_dKdz} with $C_R=1.8$, $c_2=0.1$, $u_*$, and the gradients reported in Figure \ref{Fig:gradComp}. The shaded markers indicate $z/\delta<0.1$. The one-to-one line is also shown.}
\label{Fig:Skw_predict}     
\end{figure}

Equation \ref{eqn:www_gradww} is used to compute the eddy diffusivity $K_t$ by  dividing measured $\overline{w'^3}$ with measured $\partial \sigma_w^2/\partial z$. The $K_t$ thus estimated is then normalized using $\sigma_w^2 \kappa z/u_*$ to be consistent with prior studies \citep{lopez1999wall}. The outcome is shown in Figure \ref{Fig:grad_diff}:  for $0.3 <z/\delta<0.9$, $K_t$ decays with increasing $z/\delta$; for $z/\delta<0.3$, which includes the ISL, the approach spectacularly fails even at predicting the sign of $K_t$. 
 More interesting is that this failure may be decomposed into two regions: (i) a finite $Sk_w$ associated with a zero vertical velocity variance gradient roughly in the ISL, and (ii) a negative diffusivity, mainly in the buffer region or roughness sublayer depending on the data set. The cross-over occurs when $\partial \sigma_w^2/\partial z=0$ (roughly delineated by the vertical dashed line in Figure \ref{Fig:grad_diff}). Interestingly, the work here suggests that within the ISL where $\partial\sigma_w^2/\partial z=0$, down-gradient models fail to predict a finite $\overline{w'^3}$ and a rectification based on $\overline{w'q'}$ is required in this zone. For this reason, the gradient-diffusion representation linking $\overline{w'q'}$ to $\partial K/\partial z$ in equation \ref{eq:budgetwww3} is explored in Figure \ref{Fig:wq_gradK}. Note that for $z/\delta<0.1$, data are shaded because these points are near or within the sublayer below the ISL for many of the data sets, and are in the near-wall region where there is greater uncertainty due to experimental constraints such as measurement resolution.  In this analysis, $q'$ is not measured but estimated as $q'=2 u'^2$.  By and large and for $0.08<z/\delta<0.8$, the  gradient-diffusion representation with a diffusivity based on $\kappa z u_* \phi_L(z)$ predicts well the vertical transport of energy needed to describe $Sk_w$.  It can be stated that the down-gradient model for $\overline{w'q'}$ captures the essential mechanisms needed to describe $Sk_w$ in equation \ref{eq:budgetwww4} and will be used later on to demonstrate that $Sk_w$ is dominated by this term. 
Returning to equation \ref{eq:budgetwww4}, there are two second-order velocity gradients that impact $Sk_w$.  Since the gradients are measured and the diffusivities are modeled but expected to be comparable based on the choice of $\tau'$, the significance of the gradients, normalized by $z$ and $u_*$, are first discussed in Figure \ref{Fig:gradComp}.  From this Figure, it is clear that when $z/\delta>0.1$, the normalized $\partial \sigma_u^2/\partial z$ are some 4-5 times larger in magnitude than $\partial \sigma_w^2/\partial z$.  Within the ISL, $\partial \sigma_w^2/\partial z\approx 0$ and much of the finite $Sk_w$ in the ISL is associated with $\partial \sigma_u^2/\partial z$ (as predicted by the prognostic approach).  Hence, it is conjectured that for $0.1<z/\delta<1$, the dominant term on the vertical velocity skewness is not related to $\partial \sigma_w^2/\partial z$.  That is, the vertical velocity skewness is primarily driven by
\begin{equation}
\label{eq:Skw_dKdz}
Sk_{w,t}=-\frac{2}{3}\left(1- \frac{2c_2}{C_R} \right)\frac{\kappa z u_*\phi_L(z)}{\sigma_w^3}\frac{\partial (\sigma_u^2)}{\partial z}.
\end{equation}
This conjecture is directly tested in Figure \ref{Fig:Skw_predict}, which compares measured $Sk_w$ with predictions from equation \ref{eq:Skw_dKdz} $Sk_{w,t}$ in the range of $0.1<z/\delta<1$ (measured $\sigma_u^2$ and $\sigma_w^3$ are used in these calculations). The agreement is quite acceptable given the uncertainty in assumed $\phi_L(z)$ and measured longitudinal velocity variance gradients. This finding explains why equation \ref{eqn:www_gradww} fails when not accounting for the large-scale adjustment, which is modeled here using equation \ref{eq:Skw_dKdz}. An implication of including $\phi_L$ is that in the outer layer, eddies commensurate to $\delta$ and inner layer eddies commensurate to $z$ are significant. To what degree the contribution of these eddies is evident in the experiments here is now considered using a data set where the sampling duration enables statistical convergence at the very large scales.

\subsection{Further analysis: Co-spectral results}

\begin{figure}
\centering
\noindent\includegraphics[width=20pc]{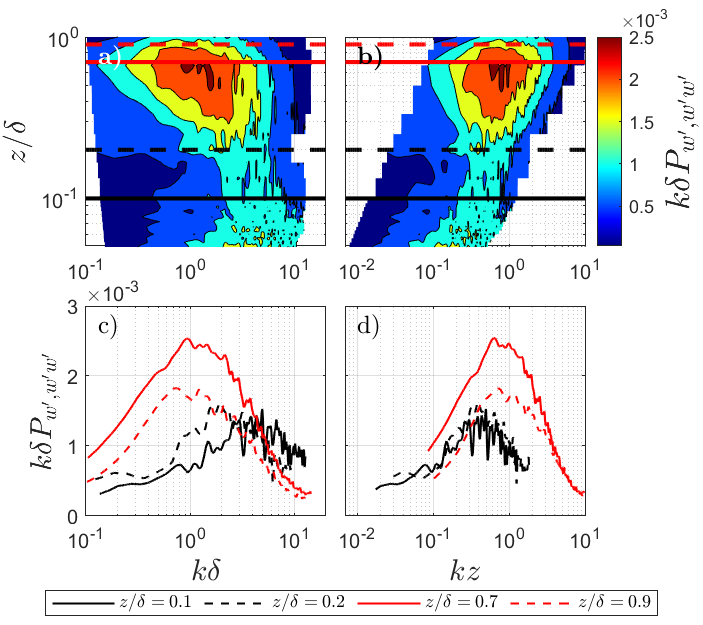}
\caption{The measured co-spectrum of $w'/\sigma_w$ and $(w'^2/\sigma_w^2)-1$ as a function of normalized wavenumber and height ($z/\delta$).  The wavenumber $k$ is normalized with $\delta$ in the left panels (a, c) and with $z$ in the right panels (b, d). The bottom panels (c,d) report the co-spectra at 4 heights: $z/\delta=0.1$ in the ISL, $z/\delta=0.2$ transitioning from the ISL to the outer layer, $z/\delta=0.7$ in the outer layer, and $z/\delta=0.9$ near the top of the boundary layer.  Those locations are also shown in the top panels as dashed horizontal lines.}
\label{Fig:cospectrum}     
\end{figure}

The co-spectrum between $w'$ and $w'^2-\overline{w'^2}$ is analyzed and a typical shape is shown in Figure \ref{Fig:cospectrum} for the longest sampling duration data set needed to resolve large and very large structures (i.e. MNS in Table \ref{tab:table1}).  Consistent with expectations from equation \ref{eq:Kw_Ku}, when $z/\delta>0.2$, scales defined by both $z$ and $\delta$ play a role as anticipated from an eddy diffusivity that scales with $z (1-z/\delta)$. At $k\delta=0.9$ the dominant length scale is $\delta$ because $z$ and $\delta$ are comparable.  Likewise, in the region where $z/\delta < 0.1$, attached eddies ($k z=0.4$) appear to contribute most to the co-spectral content of $w'$ and $w'^2-\sigma_w^2$, as expected, because $z/\delta$ no longer contributes to the eddy diffusivity.  These findings independently support the formulation in equation \ref{eq:Kw_Ku} that identifies $z$ and $\delta$ (through $\phi_L$) as the limiting scales to be accommodated in models for $Sk_w$ for the ISL and outer layer region. However, caution should be exercised in such naive binary representation of length scales as the analysis also shows that eddies up to $10\delta$ (often related to very large scale motion or VLSM) still have finite contributions on the co-spectra. 

\section{Conclusions}
The present work explores the vertical velocity skewness ($Sk_w$) in wall bounded flows covering smooth, rough, and permeable surfaces.  The exploration focused on diagnostic and prognostic models for $Sk_w$.  The following conclusions can be drawn:
\begin{itemize}
    
\item For the diagnostic models, it was shown that third order cumulant expansions for the single and joint (with $u')$ PDFs establish links between duration of updrafts $\Gamma_+$, the relative importance of ejections over sweeps to momentum transport $\Delta S_0$, and $Sk_w$.  Those derivations are statistical in nature and only offer constraints on the vertical velocity skewness values.  However, they make no contact with the Navier-Stokes equations or the physics of turbulence.

\item The fractional area/mass flux approach that is routinely used in convective boundary layer models to correct gradient-diffusion formulations was also explored (for near-neutral conditions) across many roughness values and Reynolds numbers.  This approach was used to invert for the fractional area of updrafts ($=a$) to match $\Gamma_+$.  The findings support a constant $a$ independent of $z$ in the range $0.1<z/\delta<0.3$.  When combined with third order cumulant expansion, a unique link between $a$ and $\Gamma_+$ was established. It was also shown that such a model predicts a relation between the vertical velocity flatness factor $FF_w$ and $Sk_w$ given by $FF_w=\alpha_1 (Sk_w^2+1)$ with $\alpha_1=1$ independent of $a$. Such $\alpha_1=1$ is the minimum  required to satisfy the realizability constraints.  The data here were used to examine $\alpha_1$ as a function of $z/\delta$ and it was shown $\alpha_1=3.3$ is plausible for $0.1<z/\delta<1$.  Such a value is sufficiently close to the value predicted from a quasi-Gaussian approximation ($\alpha_1=3$).

\item The prognostic approach considered the budgets for $\overline{w'}$, $\overline{w'^2}$, and $\overline{w'^3}$.  It was shown that the budget of $\overline{w'}$ yields a constant $\sigma_w^2$ independent of $z/\delta$ only when the mean pressure is hydrostatic.  This finding establishes a link between the emergence of a constant $\sigma_w^2$ with respect to $z$ expected in the ISL and models for the mean pressure.  
 
\item The budgets for $\overline{w'^2}$ and $\overline{w'^3}$, when combined together, resulted in a model for $\overline{w'^3}$ that has two contributions: a gradient diffusion contribution arising from inertia that links $\overline{w'^3}$ to $\partial \sigma_w^2/\partial z$ and a non-local contribution that links $\overline{w'^3}$ to the turbulent transport of kinetic energy $\overline{w'q'}$.  This transport term arises from return-to-isotropy considerations originally established when closing the $\overline{w'^2}$ budget using a linear Rotta scheme.  

\item The term $\overline{w'q'}$ was shown to be reasonably approximated with a down gradient model with respect to turbulent kinetic energy $K$ provided the standard eddy diffusivity $\kappa z u_*$ is multiplied by a correction function for outer layer eddies $\phi_L(z/\delta)$.  The contribution from $\overline{w'q'}$ explains much of the $Sk_w$ values in the range of $0.1<z/\delta<1$ when using $\kappa z u_* \phi_L$ and $\partial K/\partial z$.  This finding offers a bridge to the diagnostic structural models that already demonstrated $\overline{w'q'}=-(3/4) \Delta S_o (A_{SM} \sigma_u^2 \sigma_w+B_{SM} \sigma_w^3)$, where $A_{SM}=1.34$ and $B_{SM}=1.59$ are constants \citep{raupach1981conditional,heisel2020velocity}.  Thus, the vertical velocity skewness can now be written as 
\begin{align}
\label{eqn:www_gradww_conc}
     Sk_w=&-\overbrace{\frac{2}{3 C_R}\frac{\tau}{\sigma_w} \frac{\partial\overline{w'w'}}{\partial z}}^{\rm local~effects}+\\ & \underbrace{\beta_L \frac{3}{4}\Delta S_o \left(A_{SM} \frac{\sigma_u^2}{\sigma_w^2} +B_{SM}\right)}_{\rm large~scale~adjustment}\nonumber.
\end{align}
When ejections dominate momentum transport over sweeps, $\Delta S_o<0$ and $Sk_w$ remains positive even when $\partial \sigma_w^2/\partial z=0$ (i.e. negligible local effects, mean pressure is hydrostatic). This result can also be expressed as a mass flux model given the relation between $\Delta S_o$ and $\Gamma_+$ in Figure \ref{Fig:M11dSo_Gamma}, and the relation between $\Gamma_+$ and the fractional area in the mass flux approach.  This super-position of down-gradient and mass flux approaches is widely used in turbulence parameterization in climate models.  Thus, the work here offers a new perspective of how to combine local and non-local effects when modeling $Sk_w$ and shows the interconnection between diagnostic and prognostic models.

\item In canonical wall-bounded flows, large and very large scale semi-organized turbulent structures that exceed in size the boundary layer depth $\delta$ contribute significantly to turbulent kinetic energy \citep{marusic2003streamwise,townsend1980structure,banerjee2013logarithmic,katul2019anatomy} and other higher order flow statistics for $u'$ even in the ISL \citep{marusic2013logarithmic,huang2022profiles}.  To what degree these structures impact $\overline{w'^3}$ has not been resolved from earlier studies.  The work here demonstrates that such structures still have a finite contribution to the asymmetry in $w'$. To what degree their effects can be accommodated through the proposed outer layer correction $\phi_L$ to the eddy diffusivity $\kappa z u_*$ requites further exploration.
\end{itemize}

To summarize, the breadth of the results presented here provides an enhanced understanding of vertical velocity skewness that spans the flow phenomenology, governing Navier-Stokes equations, and statistical outcomes of turbulence. The asymmetry is linked to large-scale coherent turbulent structures (Figure \ref{Fig:cospectrum}), which in a simplified sense includes sweeps ad ejections. A model for skewness must therefore include a non-local adjustment to account for these large-scale eddies (equation \ref{eqn:www_gradww_gen}). The form of the adjustment can be determined prognostically from the governing budget equations (equation \ref{eq:budgetwww_cl1}) or diagnostically from statistics that quantify properties of the turbulent structures (e.g. $\Delta S_o$ and $\Gamma_+$). Regardless of the approach, the non-local adjustment is well described by the kinetic energy transport $\overline{w'q'}$ (Figure \ref{Fig:Skw_predict}) as shown here.

While these results do not address all aspects of ABL parameterizations needed in models such as CLUBB, they do offer bench-mark outcomes for the diabatic limit.  In this respect, they may be viewed as necessary but not sufficient to progress on turbulence parameterizations for asymmetry in $w'$ within climate models.  Future effort will bifurcate in two directions.  One seeks to explore the extension of these approaches (prognostic and diagnostic) to stratified flows in the atmosphere (mainly surface layer and roughness sublayer including vegetated and urban) - where buoyancy, elevated Reynolds numbers, Coriolis forces, and enhanced roughness values are expected.  The other seeks dedicated open channel flow experiments that will clarify the contribution of the many eddy sizes, including large-scale structures, on the co-spectra of $w'$- $w'^2$, $w'$ - $u'^2$, and $w'$-$w'^3$.  These co-spectra (and concomitant co-spectral peak similarities) can assist in the formulation of future prognostic models or revisions to a linear $\phi_L(z)=1-z/\delta$.  These experiments do require extended sampling duration to reliably resolve contributions of very large eddy sizes.   

\bibliography{turbbib}


\begin{acknowledgments}
EB acknowledges Politechnico di Torino (Italy) for supporting the visit to Duke University. GK acknowledges support from the U.S. National Science Foundation (NSF-AGS-2028633) and the U.S. Department of Energy (DE-SC0022072).  DP acknowledges support from Fondo europeo di sviluppo regionale (FESR) for project Bacini Ecologicamente sostenibili e sicuri, concepiti per l'adattamento ai Cambiamenti ClimAtici (BECCA) in the context of Alpi Latine COoperazione TRAnsfrontaliera (ALCOTRA) and project Nord Ovest Digitale e Sostenibile - Digital innovation toward sustainable mountain (Nodes - 4). DV and CM acknowledge European Union’s Horizon 2020 research and innovation programme under the Marie Sklodowska-Curie grant agreement No 101022685 (SHIEELD).  
\end{acknowledgments}

\end{document}